\documentstyle[prb,aps,twocolumn,floats,epsf]{revtex}

\begin{document}

\twocolumn[\hsize\textwidth\columnwidth\hsize\csname @twocolumnfalse\endcsname

\title{Interference effects in isolated Josephson
junction arrays with geometric symmetries}

\author{Dmitri A. Ivanov$^a$, Lev~B.~Ioffe$^b$,
Vadim B. Geshkenbein$^a$, and Gianni Blatter$^a$}

\address{$^a$ Theoretische Physik, ETH-H\"onggerberg, CH-8093 Z\"urich,
Switzerland \\
$^b$ Center for Materials Theory, Physics Department, Rutgers University, 
Piscataway, NJ 08854 USA}

\date{February 9, 2001}

\maketitle

\begin{abstract}
  
  As the size of a Josephson junction is reduced, charging effects become
  important and the superconducting phase across the link turns into a
  periodic quantum variable. Isolated Josephson junction arrays are described
  in terms of such periodic quantum variables and thus exhibit pronounced
  quantum interference effects arising from paths with different winding
  numbers (Aharonov--Casher effects). These interference effects have strong
  implications for the excitation spectrum of the array which are relevant in
  applications of superconducting junction arrays for quantum computing. The
  interference effects are most pronounced in arrays composed of identical
  junctions and possessing geometric symmetries; they may be controlled by
  either external gate potentials or by adding/removing charge to/from the
  array. Here we consider a loop of $N$ identical junctions encircling one
  half superconducting quantum of magnetic flux. In this system, the ground
  state is found to be non-degenerate if the total number of Cooper pairs on
  the array is divisible by $N$, and doubly degenerate otherwise (after the
  stray charges are compensated by the gate voltages).

\end{abstract}


]

\section{Introduction}

Josephson junction arrays are excellent tools for exploring quantum-mechanical
behavior in a wide range of parameter space~\cite{JJA}.  The charge and phase
on each island provide a set of conjugate quantum variables, allowing for dual
descriptions of the array either in terms of charges (Cooper pairs) hopping
between the islands or in terms of vortices hopping between the plaquets of
the array~\cite{Blanter}. This opens the door for the set up and manipulation
of interesting quantum interference effects in Josephson junction arrays: in a
magnetic field, charges pick up additional Aharonov--Bohm phases and hence the
properties of the array depend on the field strength. In the dual language,
vortices moving around islands gain phases proportional to the average charges
on the islands (Aharonov--Casher phases)~\cite{Aharonov-Casher,Mooij3}.  These
features have recently been used in various proposals for solid state
implementations of qubits for quantum computing, based on either the
charge~\cite{Schoen} or phase~\cite{Levitov,Blatter,Blatter-2} degree of
freedom in Josephson junction arrays.

In this paper we study an interference effect in electrically isolated
Josephson junction arrays which renders the ground state and excitation
spectrum sensitive to the total charge on the array. This effect combines the
dual descriptions in terms of charge or phase: {\it i)} in the limit where the
charging energy $E_C$ is much larger than the Josephson energy $E_J$, the
fluctuations of charge on the islands are small and the system is equivalent
to strongly repulsive bosons (Cooper pairs) hopping between islands; the total
charge then determines the number of such bosons and hence the structure of
the spectrum. {\it ii)} In the opposite limit of large Josephson energy $E_J$,
the phase fluctuations are small and the spectrum is determined by phase
tunneling between classical minima of the Josephson energy; the total charge
on the array then determines the interference between different tunneling
trajectories via the Aharonov--Casher effect. These interference effects,
where the ground state and the excitations may change degeneracies depending
on the total charge on the array, are most pronounced in small arrays with
geometric symmetries; for the symmetric loops considered in this paper they
coincide in the charge- and phase-dominated limits and persist at arbitrary
ratio of the Josephson- to charging energy. In mathematical terms, the total
charge on the array enters its symmetry group producing a central
extension~\cite{central-extension}. The irreducible representations are then
classified by the total charge, and their dimensions (and associated level
degeneracies) depend on the charge on the array.

\begin{figure}[!b]
\centerline{\epsfxsize = 6.0cm \epsfbox{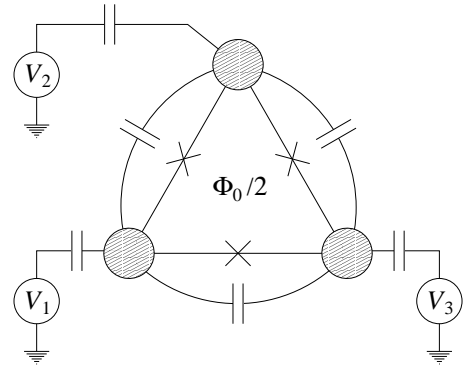}} \narrowtext\hspace{0.5cm}
\caption{A superconducting phase qubit consists of a small inductance loop made
  from $N$ superconducting islands connected by Josephson junctions and
  capacitively coupled to each other and to the ground (the case $N=3$ is
  shown here). The loop is placed in an external magnetic field producing one
  half superconducting flux quantum through the loop.}
\label{qubit}
\end{figure}

In experimental realizations of Josephson junction arrays, the charges and
potentials on the islands are affected by a number of external factors such as
differences in island size, background charges, charges localized at
impurities, etc. All these factors produce random charge offsets on the
islands which distort the quantum interference pattern (see e.g.\ the
discussion of offset charges in Ref.~\onlinecite{JJA}). In order to observe
the quantum interference effects predicted in this paper the potentials of the
islands should be adjustable via external capacitive gates. The offset charges
should then be compensated by appropriate tuning of the gate voltages; such a
procedure was successfully carried out in a transport measurement involving
two islands~\cite{Mooij1} and should also be possible for the system
considered in this paper. We also assume the absence of quasiparticles on the
islands. Firstly, this requires the temperature to be sufficiently low to
eliminate thermal quasiparticles. Secondly, the total number of electrons on
the array must be even in order to exclude unpaired charge due to the parity
effect.

In the following we study the simplest setup exhibiting these interference
effects, a closed low-inductance loop with $N\ge 3$ identical Josephson
junctions and pierced by one-half superconducting flux quantum
(Fig.~\ref{qubit}). Such loops have been proposed as possible realizations for
superconducting phase quantum bits~\cite{Levitov,Blatter,Blatter-2} and a
successful quantum superposition of the two qubit states has been reported
recently~\cite{Mooij2}.  While in the original qubit design the loop has been
made asymmetric in order to suppress the interference of different tunneling
paths~\cite{Levitov,Blatter-2}, it is this interference that we study in the
present paper where we assume the islands and contacts to be nearly identical
(the required precision will be specified at the end of the paper).  The
ground state degeneracy of such a symmetric Josephson-junction loop is found
to depend periodically on the total charge $Q$ with period $2eN$, $N$ is the
number of islands in the loop. In particular, in the phase-dominated regime
$E_J\gg E_C$ the Josephson energy of the fully frustrated loop exhibits two
equivalent minima describing states with currents circulating in opposite
directions. Tunneling between these minima produces a splitting $\Delta$
between these levels which is determined by the charges $q_i$ (in units of
$2e$) induced on the islands,
\begin{eqnarray}
\label{splitting}
\Delta&=&\Delta_0 \left| 1+e^{2\pi iq_1}+e^{2\pi i(q_1+q_2)}+\dots \right.
\\
&&\qquad\qquad\qquad\qquad\left. +e^{2\pi i(q_1+\dots+q_{N-1})} \right|,
\nonumber
\end{eqnarray}
and which vanishes for $(Q/2e)\neq 0~({\rm mod}~N)$ (the sum of all induced
charges $q_i$ equals $Q$, and the expression (\ref{splitting}) is symmetric
under circular permutation of the islands). This result is the consequence of
the interference of different tunneling paths connecting the two minima where
the relative phases of the tunneling amplitudes depend on the charges $q_i$ (a
similar effect was predicted for the S-S-S\- double junction in
Ref.~\onlinecite{Ivanov}). The induced charges $q_i$ may be tuned by either
gate voltages (redistributing the existing charge) or by adding/removing extra
charge to/from the array. E.g. suppose that all charges $q_i$ have been set to
zero by fine tuning the gate voltages, thus maximizing the splitting $\Delta$.
Then adding (or removing) a charge $Q$ at fixed relative potentials adds a
value $Q/2eN$ to each induced charge $q_i$ and closes the gap $\Delta$: the
splitting is present when the charge $Q/2e$ is divisible by $N$ and absent
otherwise. A similar result applies for the opposite limit $E_C\gg E_J$ where
the ground state is non-degenerate with the next excited state an energy $E_C$
away if $Q/2e$ is divisible by $N$ (`insulating' state), while it is
degenerate with an excitation gap of order $E_J$ if 
$Q/2e \neq 0~({\rm mod}~N)$ (`metallic' state). Indeed, this $Q$-dependence 
of the ground state degeneracy will be explained using symmetry 
considerations valid in the entire range of couplings $E_J/E_C$.

The paper is organized as follows. In Section II, we define the
model and prove the periodic dependence of the excitation spectrum
on the total charge $Q$, for a symmetric loop. In Sections III
and IV, we treat the limits $E_J\gg E_C$ and $E_J\ll E_C$,
respectively. Section V contains the analysis of the symmetries
of the loop. Finally, in Section VI we discuss the physical
requirements for observing the charge-dependent interference
effects.

\section{Model and periodic $Q$-dependence of the spectrum}

The Hamiltonian describing the qubit in Fig.~\ref{qubit} takes the form
\begin{eqnarray}
H&=&{1\over2} \sum_{ij} Q_i (C^{-1})_{ij} Q_j + \sum_i V_i Q_i \nonumber \\
&&\qquad\qquad + \sum_i U_i(\varphi_{i+1}-\varphi_i-a_{i,i+1}).
\label{Ham1}
\end{eqnarray}
Here, $\varphi_i$ are the phases on the islands, $Q_i \equiv -i\partial/
\partial\varphi_i$ are the charge operators conjugate to $\varphi_i$ (we
measure $Q$ in units of $2e$ from now on), $(C^{-1})_{ij}$ is the inverse
capacitance matrix, $V_i$ are the gate voltages applied to the islands, $U_i$
are the Josephson energies of the junctions, and $a_{i,i+1}$ is the
electromagnetic vector potential induced by the external magnetic field ($i+1$
in the indices should be understood modulo $N$). This Hamiltonian acts on the
wave functions $\Psi(\varphi_1, \dots, \varphi_N)$ which are periodic in all
their variables,
\begin{eqnarray}
\label{boundary-conditions}
&& \Psi(\varphi_1, \dots, \varphi_i + 2\pi, \dots, \varphi_N) \\
&&\qquad\qquad\qquad = \Psi(\varphi_1, \dots, \varphi_i, \dots, \varphi_N).
\nonumber
\end{eqnarray}
In general, the boundary conditions (\ref{boundary-conditions}) may contain
arbitrary phase shifts $e^{i\theta_i}$ incorporating the effect of background
charges. They can be manipulated by the gate voltages $V_i$ and we assume them
to vanish through appropriate fine tuning\cite{Mooij1} (this is equivalent to
adjusting the zero positions of the gate voltages).  We assume that this
tuning is performed once in the beginning of the experiment and later measure
the gate voltages relative to these reference values.

Since the potential term in the Hamiltonian (\ref{Ham1}) contains only phase
differences $\varphi_i-\varphi_j$, it has the symmetry of rotating all the
phases by the same angle: $\Psi(\varphi_1, \dots, \varphi_N) \mapsto
\Psi(\varphi_1+\delta\varphi, \dots, \varphi_N+\delta\varphi)$.  Equivalently,
the total charge $Q=\sum_i Q_i$ is conserved (i.e., the charge $Q$ commutes
with the Hamiltonian $H$) and we may project the Hilbert space onto the
subspace with a given total charge $Q$ before diagonalization, implying the
transformation rule
\begin{equation}
\label{overall-phase-shift}
\Psi(\varphi_1+\delta\varphi, \dots, \varphi_N+\delta\varphi)
=e^{i\delta\varphi Q} \Psi(\varphi_1, \dots, \varphi_N)
\end{equation}
for the simultaneous rotation of all the phases by $\delta\varphi$.  From the
periodicity of $\Psi$, it immediately follows that $Q$ is integer, i.e., the
total charge must be a multiple of $2e$ (this is in fact an implicit
assumption when writing the Hamiltonian (\ref{Ham1}) in terms of phases only).
In the following we shall discuss the symmetric loop, postponing the effects
of asymmetry till the end of the paper.  Here, by symmetry we mean
$(C^{-1})_{ij}=(C^{-1})_{i+k,j+k}$ for the Coulomb term and equality of all
Josephson terms $U_i(\varphi)$. This implies that the loop is invariant under
circular permutation of the islands. Also, if the flux through the loop is
exactly one half flux quantum, the loop is invariant under `flips' changing
the direction of the current.

The excitation spectrum of the symmetric loop 
periodically depends on the total charge
$Q$ with the period $N$ (up to overall shifts): the unitary operator
\begin{equation}
U: \Psi \mapsto \Psi e^{i(\varphi_1+\dots+\varphi_N)}
\label{U}
\end{equation}
increases the charge on all the islands by one, $U^{-1} Q_i U = Q_i +1$, and
therefore the sector with total charge $Q$ maps onto the one with total charge
$Q+N$. On the other hand,
\[
U^{-1} H U = H + \sum_i (C^{-1})_{ij} Q_j + {1\over2} \sum_{ij} (C^{-1})_{ij} +
\sum_i V_i.
\]
If all the islands are equivalent, then the sum $\sum_i (C^{-1})_{ij}$ is
independent of $j$ and
\begin{equation}
U^{-1} H U = H + {N\over C} \left(Q+{N\over2}\right) + \sum_i V_i,
\label{energy-shift}
\end{equation}
where $C=N\big[\sum_i (C^{-1})_{ij}\big]^{-1} = \sum_{ij} C_{ij}$ is the total
capacitance of the loop. Thus the operator $U$ maps the eigenfunctions of the
Hamiltonian in the sector with charge $Q$ onto eigenfunctions in the sector
with charge $Q+N$ shifting them in energy by a constant as given by
(\ref{energy-shift}), thus proving our statement about the periodic
$Q$-dependence of the excitation spectrum.

\section{Level splitting in the $E_J\gg E_C$ limit}

In the phase-dominated regime with $E_J \gg E_C$ the low-energy states of the
Josephson junction loop are determined by the classical minima of the
Josephson energy as corrected by weak quantum tunneling (due to the finite
charging energy).  Here, we consider a symmetric loop pierced by half a
quantum of magnetic flux, $\sum_i a_{i,i+1}=\pi$, and we choose to work in a
gauge with $a_{i,i+1}=\pi/N$ for all $i$. In the limit $E_J\gg E_C$ the only
constraint on the potential $U_i(\varphi)$ is the double degeneracy of the
total Josephson energy as a function of the phases $\varphi_i$ (e.g., this
requirement is satisfied for tunneling junctions with $U_i(\varphi)= -E_J
\cos\varphi$ and $N \geq 3$).  The two potential minima are determined by the
phase configurations $\varphi_i=0$ and $\varphi_i=(2\pi/N)i$ (and all
configurations obtained from these two by the continuous symmetry $\varphi_i
\mapsto \varphi_i+\delta\varphi$). These two minima involve different
directions of the Josephson current, circulating the loop clockwise or
counter-clockwise. The continuous symmetry $\varphi_i \mapsto \varphi_i +
\delta\varphi$, $i=1,\dots,N$ implies the quantization of the total charge $Q$
as discussed above.

\begin{figure}[!t]
\centerline{\epsfxsize = 7.0cm \epsfbox{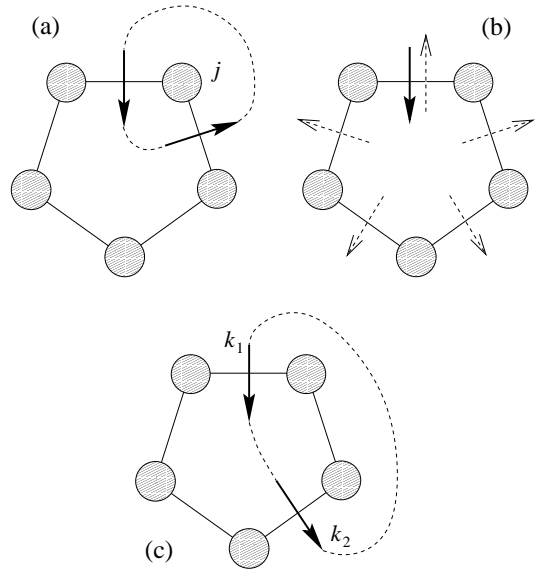}} \narrowtext\hspace{0.5cm}
\caption{(a) One flux quantum moving around the $j$-th island; (b) One of the
  $N$ possible tunneling trajectories: one flux quantum enters the loop
  through one of the junctions and leaves divided uniformly between all
  junctions; (c) The phase difference between two tunneling trajectories with
  flux entry through the junctions $k_1$ and $k_2$ is given by the
  Aharonov--Casher phase arising from one flux unit circling the (charged)
  islands in between. Note that closing the flux trajectory in (c) on the
  opposite side produces the same phase because $Q$ is an integer.}
\label{fluxes}
\end{figure}

In the following it will be convenient to incorporate the voltage terms $V_i
Q_i$ into the quadratic kinetic term via a shift $Q_i \mapsto Q_i-q_i$, where
\begin{equation}
q_i=\sum_j C_{ij} \left(V_i - {\sum_k V_k \over N}\right) + {Q\over N}
\end{equation}
are the mean charges induced on the islands (we also include the
$i$-independent contribution from the total charge $Q$ for further
convenience). This gauge transformation eliminates the $V_i Q_i$ terms in the
Hamiltonian (and adds a constant energy shift which we disregard). Under this
transformation the wave functions pick up additional phases $\Psi \mapsto
\tilde\Psi = \exp(-\sum_i q_i\varphi_i)\Psi$; the new wave functions
$\tilde\Psi$ satisfy the twisted boundary conditions
\begin{eqnarray}
  \label{twisted_bc}
  && \tilde\Psi(\varphi_1, \dots, \varphi_j + 2\pi, \dots, \varphi_N) \\
  && \qquad\qquad = \exp (-2\pi i q_j)
  \tilde\Psi(\varphi_1, \dots, \varphi_j, \dots, \varphi_N) \nonumber
\end{eqnarray}
and the invariance under the continuous symmetry $\varphi_i \mapsto \varphi_i
+ \delta\varphi$ takes the form (c.f.~(\ref{overall-phase-shift})),
\begin{equation}
  \label{invariance}
  \tilde\Psi(\varphi_1+\delta\varphi, \dots,
  \varphi_N+\delta\varphi) = \tilde\Psi(\varphi_1, \dots, \varphi_N).
\end{equation}
We introduce the new variables $\phi_+=(\varphi_1+\dots+\varphi_N)/N$
(conjugate to $Q$) and the gauge invariant phase drops over the junctions
$\phi_i=\varphi_{i+1}-\varphi_i-a_{i,i+1}$ (out of $N$ variables $\phi_i$,
only $N-1$ are independent, since $\sum_i \phi_i = -\pi$). The Hamiltonian
then decouples into two, one involving only $\phi_+$ and the other only the
phase drops $\phi_i$. The continuous symmetry (\ref{invariance}) implies that
$\tilde\Psi$ is independent of $\phi_+$, i.e., $\tilde\Psi \equiv
\tilde\Psi(\phi_1,\dots, \phi_N)$. The boundary conditions on $\tilde\Psi$
derived from (\ref{boundary-conditions}) then take the form
\begin{eqnarray}
  \label{bc}
  && \tilde\Psi(\phi_1,\dots,\phi_j+2\pi, \phi_{j+1}-2\pi,\dots, \phi_N) \\
  && \qquad\qquad = \exp (2\pi i q_j)
  \tilde\Psi(\phi_1, \dots, \phi_j, \phi_{j+1}, \dots, \phi_N)\nonumber
\end{eqnarray}
and can be given a simple physical interpretation (see Fig.~\ref{fluxes}(a)):
on moving a flux $2\pi$ (one flux quantum) around the $j$-th island the wave
function gains a phase $2\pi q_j$.

We are now prepared for the calculation of the tunneling amplitude connecting
the two semiclassical minima.  Due to the equivalence of phase drops $\phi_i$
differing by a multiple of $2\pi$, the coordinate space is a
$(N-1)$-dimensional torus.  After unfolding the torus onto the plane, each
point on the torus is represented by a lattice of points on the plane. We
choose the point $\{\phi_i=-\pi/N\}$ as the representative of one of the two
minima. The $N$ neighboring minima with coordinates $\{\phi_i=\pi/N,~i\neq
k;~\phi_k=\pi/N-2\pi\}$, $k=1,\dots,N$ then correspond to the other minimum.
Accordingly, there are $N$ different optimal tunneling trajectories connecting
the two minima and we have to add up the various amplitudes coherently in
order to find the hopping amplitude.  Inspection of the particular phase
changes along each of these trajectories shows that they may be thought of as
describing a flux $2\pi$ entering the loop through the junction $k$ and
leaving the loop equally distributed (i.e., with a fraction $2\pi/N$ of flux
per junction) among all the junctions, see Fig.~\ref{fluxes}(b).  These
trajectories exhibit different phase windings and thus are topologically
different. The relative phases between these trajectories may then be read off
the boundary conditions for the wave function $\tilde\Psi$; e.g., the phase
difference between the tunneling amplitudes for the trajectories $k_1$ and
$k_2$ equals the phase difference of the wave function $\tilde\Psi$ at the
points $\{\phi_i=\pi/N,~ i \neq k_1;~\phi_{k_1}=\pi/N-2\pi\}$ and
$\{\phi_i=\pi/N~i \neq k_2;~ \phi_{k_2}=\pi/N-2\pi\}$. This phase difference
is $2\pi(q_{k_1+1}+q_{k_1+2}+\dots+q_{k_2})$ and may be interpreted as the
phase generated by the unit flux entering the loop through junction $k_1$ and
leaving through junction $k_2$ (see Fig.~\ref{fluxes}(c)).  From this argument
immediately follows the result (\ref{splitting}) for the level splitting (the
level splitting in the double-well problem is proportional to the absolute
value of the tunneling amplitude).

\section{Level splitting in the $E_J \ll E_C$ limit}

In the charge-dominated limit $E_J \ll E_C$ it is convenient to work in the
charge representation~\cite{Mooij-charge}, where the operators $Q_i$ are
diagonal and the Josephson term in the Hamiltonian (\ref{Ham1}) takes the form
(we restrict the discussion to tunneling junctions with $U_i(\varphi) = - E_J
\cos(\varphi)$)
\begin{equation}
H_J=-{E_J\over 2}\sum_i \left(L_i^+ L_{i+1}^- e^{i\pi/n} + {\rm h.c.}
\right),
\label{HJ}
\end{equation}
with $L_i^+$ ($L_i^-$) the charge raising (lowering) operators on the $i$-th
island, $L_i^\pm|Q_i\rangle=|Q_i\pm 1\rangle$. The periodicity in $Q$ allows
us to restrict our analysis to the $N$ charge sectors $Q=0,\dots,N-1$. Below
we first neglect the coupling $E_J$ and find the ground states of the Coulomb
part of the Hamiltonian; hopping between the islands is then perturbatively
included in a second step.

\begin{figure}[!b]
\centerline{\epsfxsize = 7.0cm \epsfbox{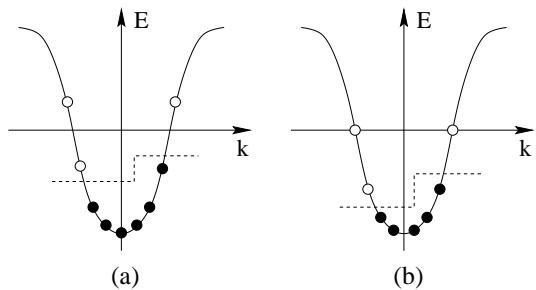}} \narrowtext\hspace{0.5cm}
\caption{The tight-binding spectrum on the circle with periodic (a) and
  anti-periodic (b) boundary conditions applying for even and odd $Q$,
  respectively. The solid circles denote filled states at the bottom
  of the band.  For $Q$ not divisible by $N$, the ground state is
  doubly degenerate. $Q$ divisible by $N$ corresponds to an empty
  (or filled) band.}
\label{tight-binding}
\end{figure}

We start with a diagonal matrix $(C^{-1})_{ij}$ and ignore the capacitive
coupling of the junctions, i.e., the islands are coupled only to the ground
but not to each other. In that case, the ground state of the Coulomb part of
the Hamiltonian is $C_N^Q$-fold degenerate ($C_N^Q=N!/Q!(N-Q)!$ 
enumerates the
number of ways to distribute $Q$ particles among $N$ sites without double
occupancy).  Second, we
solve the Hamiltonian (\ref{HJ}) projected onto this $C_N^Q$-dimensional
subspace in order to find the level splitting at finite $E_J$. This is easily
done by observing the equivalence of this problem to the tight-binding model
for hard-core bosons on the circle with $N$ sites.  Mapping to free fermions
and taking into account the flux through the loop and the boundary conditions
on the circle, one finds that the projected $H_J$ describes a tight binding
model for $Q$ free fermions on a circle with $N$ sites and {\it periodic}
boundary conditions for even $Q$, while {\it anti-periodic} boundary
conditions apply for odd $Q$ (Fig.~\ref{tight-binding}). 
This implies that the ground state is
non-degenerate with a gap of order $E_C$ if $Q$ is divisible by $N$ (the
`insulating state' with an empty band) and doubly degenerate (with the lowest
excitation at an energy of the order of the intra-band level spacing $E_J/N$
above the degenerate ground state) otherwise.

Turning on the junction capacitance makes the particles (Cooper pairs) repel
each other and they tend to arrange in configurations with maximal separation
between them. The number of such configurations is generally less (or equal)
than $C_N^Q$. We conjecture that even in this case the ground state level is
degenerate unless $Q$ is divisible by $N$. We do not have a rigorous proof of
this statement but have verified it for $N=3,4,5$ (in fact, for $N=3$ the
off-diagonal elements of the matrix $(C^{-1})_{ij}$ may be incorporated in a
constant term $\propto Q^2$ and do not change the properties of the system).
The degeneracy may, of course, be explicitly verified for any given $N$ and
$Q$.

\section{Symmetry analysis}

\begin{figure}[!b]
\centerline{\epsfxsize = 7.0cm \epsfbox{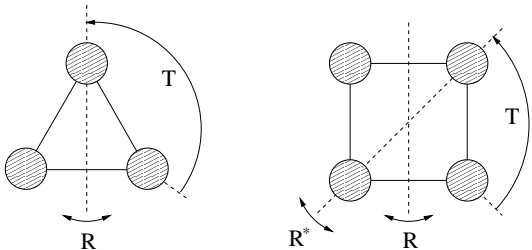}} \narrowtext\hspace{0.5cm}
\caption{Geometric symmetries of the Josephson junction loop for odd and even
  $N$ involving translations $T$ and reflections $R$. For even $N$, the two
  classes of reflections are related by $R^*=TR$.}
\label{symmetries}
\end{figure}

In the previous sections, we have seen that the ground-state degeneracy
depends on the divisibility of the total charge $Q$ by the number of
islands $N$ and coincides in the two limits $E_J\ll E_C$ and
$E_J\gg E_C$. In this section we show that this degeneracy is a
consequence of the geometric symmetries of the loop and remains
exact beyond the perturbation theory around these limiting cases.
We shall classify the irreducible representations of the symmetry
group of the loop, and identify the representations corresponding
to the ground state. In both limits, the ground states correspond
to the same representation, which implies that the degeneracy is
also preserved in the intermediate parameter range, unless level-crossing
occurs. We verify the absence of level-crossing in the
several simplest cases numerically by exact diagonalization.

The level degeneracies for general values of $E_J/E_C$ are determined by the
symmetry group of the Josephson-junction loop. 
The geometric symmetries are described by the dihedral group $D_N$
consisting of cyclic permutations $T$ of the islands (preserving the loop
orientation; these are equivalent to a $N$-fold rotation axis) and of
reflections $R$ about diameters of the loop drawn through islands or through
junctions, see Fig.~\ref{symmetries}. The reflections $R$ 
reverse the flux through
the loop and belong to its symmetry group for flux zero or half-quantum. The
dihedral group $D_N$ involves $2N$ symmetry operations; it may be
characterized with the defining relations
\begin{eqnarray}
  T^N &=& 1, \label{def_1a} \\
  R^2 &=& 1, \label{def_1b} \\
  (TR)^2 &=& 1. \label{def_1c}
\end{eqnarray}

These relations are obeyed by the operators representing $R$ and $T$
at zero magnetic flux through the loop. On the other hand, for the 
case of half a flux quantum piercing the loop the symmetry operators 
preserving the Hamiltonian need to be supplemented by additional gauge 
transformations. As a consequence, the above relations are modified,
with the appearance of additional phase shifts. Explicitly,
the operators corresponding to rotations $T$ and reflections $R$ 
have the form
\begin{eqnarray}
  T \Psi(\varphi_1, \dots, \varphi_N)
  &=& \Psi(\varphi_2, \dots, \varphi_N, \varphi_1), \label{TPsi} \\
  R \Psi(\varphi_1, \dots, \varphi_N)
  &=& \Psi(\varphi_N, \varphi_{N-1}+{2\pi}/{N},
     \dots, \label{RPsi}\\
  &&\qquad\qquad\varphi_1+{2\pi}(N-1)/N),\nonumber
\end{eqnarray}
where the shifts $(2\pi/N) k$ of the phase $\varphi_{N-k}$ compensate for the
external vector potential in the Hamiltonian (\ref{Ham1}) and guarantee its
invariance. This modifies the relation (\ref{def_1b}) producing an additional
phase shift:
\begin{equation}
  R^2 = \exp(-2\pi i Q/N). \label{def_2b}
\end{equation}

The simplest way to make use of symmetry arguments is for the
case $N$ even and $Q$ odd: we show that all states are 
degenerate in this case. Let us assume the opposite, i.e., that the
eigenstate $|\Psi\rangle$ is non-degenerate. Acting with $T$
and $R$ on $|\Psi\rangle$ we reproduce the state up to phases
$t$ and $r$ which satisfy the relations $t^N = (tr)^2 = 1$ and
$r^2 = \exp(-2\pi i Q/N)$; this set of equations is inconsistent 
for $N$ even and $Q$ odd, hence all levels indeed are degenerate. 

To determine the level degeneracy and to treat the case of
arbitrary $N$ and $Q$, we classify the representations of the
symmetry group.
To take into account the phase shift in (\ref{def_2b}), 
we include such phase shifts as new central elements $\{Z^n|_{n=1}^N\}$
in the group.  The resulting set of defining relations is
\begin{eqnarray}
&& Z^N=T^N=(TR)^2=1, \quad
R^2=Z^{-1}, \nonumber\\
&& ZR=RZ, \quad
ZT=TZ
\label{def_ZD}
\end{eqnarray}
(the last two relations are equivalent to the statement that $Z$ commutes with
all group elements, i.e., it is a central element).  The set of relations
(\ref{def_ZD}) defines a {\it central extension} of the group $D_N$ which we
further denote $ZD_N$. Extending $D_N$ to $ZD_N$ by the central element $Z$
enlarges the number of elements in the group by a factor of $N$, and the
number of elements of $ZD_N$ is $2N^2$.

In any particular irreducible representation, $Z$ is represented by a number
(a $N$-fold root of unity, since $Z^N=1$):
\begin{equation}
Z=\exp(2\pi i Q/N)
\end{equation}
This formula establishes the relation between the representations of $ZD_N$
and the total charge $Q$ on the array.

\begin{table}
\caption{Character table of the dihedral group $D_N$ for $N$~odd:
the two one-dimensional representations are 
denoted by $D^{\scriptscriptstyle(\pm)}$; the
$(N-1)/2$ two-dimensional representations 
$D^{\scriptscriptstyle(\mu)}$ involve the roots
$t_\mu=\exp(2\pi i\mu/N)$, with $\mu=1,\dots,(N{-}1)/2$.}
  \vskip 3 pt
  \begin{tabular}{l|ccc}
          & \{E\} & $\{T^\nu,T^{-\nu}\}_{\nu=1}^{(N-1)/2}$ 
          & $\{T^{m}R\,|_{m=0}^{N-1}\}$\\
            \noalign{\vskip 3 pt}
            \hline\noalign{\vskip 3 pt}
  $D^{\scriptscriptstyle(\pm)}$ & 1 & 1 & $\pm 1$\\
            \noalign{\vskip 3 pt}
  $D^{\scriptscriptstyle(\mu)}$ & 2 & $t_\mu^\nu+t_\mu^{-\nu}$ & 0 \\
  \end{tabular}
\label{table1}
\end{table}
\begin{table}
\caption{Character table of $ZD_N$ for $N$~odd. The representations
are classified by the charge $Q$ which relates to the values of the
central element $Z = \exp(2\pi i Q/N)$.
The $2N$ one-dimensional representations 
are denoted by $D^{\scriptscriptstyle(\pm,Q)}$; the $N(N-1)/2$
two-dimensional representations $D^{\scriptscriptstyle(\mu,Q)}$ 
involve the roots $t_\mu=\exp(2\pi i\mu/N)$. The indices $n$ and 
$\mu$ span the integers $n=0,\dots,N-1$ and $\mu=1,\dots,(N-1)/2$, 
respectively.}
  \vskip 3 pt
  \begin{tabular}{l | c c c }
          & $\{Z^n\}$ & $\{Z^nT^\nu,Z^{n+\nu}T^{-\nu}\}_{\nu=1}^{(N{-}1)/2}$
          & $\{Z^{n-m}T^{2m}R\,|_{m=0}^{N-1}\}$\\
            \noalign{\vskip 3 pt}
            \hline\noalign{\vskip 3 pt}
  $D^{\scriptscriptstyle(\pm,Q)}$ & $Z^n$ & $Z^{n+\nu/2}$ & $\pm Z^{n-1/2}$\\
            \noalign{\vskip 3 pt}
  $D^{\scriptscriptstyle(\mu,Q)}$ & $2Z^n$ & 
       $Z^{n+\nu/2}(t_\mu^\nu+t_\mu^{-\nu})$ & 0 \\
\end{tabular}
\label{table2}
\end{table}

Here we should mention that central extensions of symmetry groups are common
in quantum mechanics. Indeed, the overall phase of the wave function has no
physical meaning. Therefore the operator representing the product of two
symmetry operations must equal the product of the two operators representing
each of these operations only up to an overall phase factor. In other words,
quantum mechanics admits not only linear representations of the symmetry
group, but a more general class of {\it projective}
representations~\cite{central-extension}.  At the same time any projective
representation of a group corresponds to a linear representation of a central
extension of this group.  Physical examples are numerous, including
half-integer spin (projectively representing the rotation group), magnetic
translations (projectively representing geometric translations), and anyons
(projectively representing the braid group). An example resembling the
analysis in the present paper is given by Kalatsky and
Pokrovsky\cite{Pokrovsky} in their discussion of the spectrum of large spins
in external crystal electric fields; the degeneracies then are described in
terms of projective representations of a finite symmetry group which depend on
the spin assuming integer or half-integer values.

\begin{figure}[!t]
\centerline{\epsfxsize = 7.0cm \epsfbox{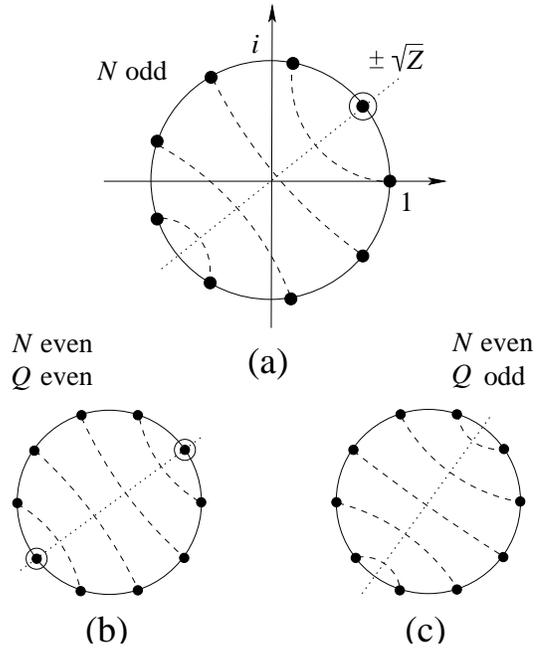}}
\narrowtext\hspace{0.5cm}
\caption{The irreducible representations of $ZD_N$ may be
  distinguished by the eigenvalues of the operator $T$. These eigenvalues must
  belong to the set of $N$-fold roots of unity $\{\exp(2\pi i n/N)\}$ (marked
  by filled dots on the unit circle). The dotted line is drawn to intersect
  the unit circle at points $\pm\sqrt{Z}$.  {\bf (a)} $N$ odd: exactly one of
  the square roots $\pm\sqrt{Z}$ is a $N$-fold root of unity. It gives the
  eigenvalue of $T$ involved in the one-dimensional representations
  $D^{\scriptscriptstyle(\pm,Q)}$ (marked by a small circle around the filled
  dot). The rest of allowed eigenvalues combine in pairs to form
  two-dimensional representations $D^{\scriptscriptstyle(\mu,Q)}$ (the
  corresponding pairs of dots are connected by dashed lines).  {\bf (b)} and
  {\bf (c)}: the corresponding illustrations for even $N$ and even $Q$, and
  for even $N$ and odd $Q$ respectively.}
\label{reps}
\end{figure}

\begin{table*}
\caption{Character table of the dihedral group $D_N$ for $N$~even:
the four one-dimensional representations are 
denoted by $D^{\scriptscriptstyle(\pm)}_\pm$; the
$N/2-1$ two-dimensional representations 
$D^{\scriptscriptstyle(\mu)}$ involve the roots
$t_\mu=\exp(2\pi i\mu/N)$ with $\mu=1,\dots,N/2{-}1$.}
  \vskip 3 pt
  \begin{tabular}{l|ccccc}
          & \{E\} & $\{T^\nu,T^{-\nu}\}_{\nu=1}^{N/2{-}1}$
          & $\{T^{N/2}\}$ & $\{T^{2m}R\,|_{m=0}^{N/2{-}1}\}$
          & $\{T^{2m+1}R\,|_{m=0}^{N/2{-}1}\}$ \\
            \noalign{\vskip 3 pt}
            \hline\noalign{\vskip 3 pt}
  $D^{\scriptscriptstyle(\pm)}_+$ & 1 & 1 & 1 & $\pm 1$ & $\pm 1$\\
            \noalign{\vskip 3 pt}
  $D^{\scriptscriptstyle(\pm)}_-$ & 1 & $(-1)^\nu$ & $(-1)^{N/2}$ & 
                                                $\pm 1$ & $\mp 1$\\
            \noalign{\vskip 3 pt}
  $D^{\scriptscriptstyle(\mu)}$ & 2 & $t_\mu^\nu+t_\mu^{-\nu}$ &
                           $2(-1)^\mu$ & 0 & 0 \\
  \end{tabular}
\label{table3}
\end{table*}
\begin{table*}
\caption{Character table of $ZD_N$ for $N$~even. The representations
are classified by the charge $Q$ which relates to the values of the
central element $Z = \exp(2\pi i Q/N)$.
The $2N$ one-dimensional representations $D^{\scriptscriptstyle(\pm,Q)}_\pm$ 
exist for even $Q$; the $N(N-1)/2$ two-dimensional representations 
denoted by $D^{\scriptscriptstyle(\mu,Q)}$ involve 
the roots $t_\mu=\exp(2\pi i\mu/N)$. The indices $n$, $m$, and $\mu$ span the 
integers $n=0,\dots,N-1$, $m = 0,\dots,N/2-1$, and $\mu=1,\dots,N/2-1$, 
respectively; the index $\bar{\mu}$ runs over the half-integers 
$\bar{\mu} = 1/2,3/2,\dots,(N-1)/2$.}
  \vskip 3 pt
  \begin{tabular}{l|ccccc}
          & $\{Z^n\}$ & $\{Z^nT^\nu,Z^{n+\nu}T^{-\nu}\}_{\nu=1}^{N/2{-}1}$ &
          $\{Z^m T^{N/2},Z^{m{+}N/2}T^{N/2}\}$ &
          $\{Z^{m-n}T^{2n}R\,|_{n=0}^{N-1}\}$ &
          $\{Z^{m-n}T^{2n+1}R\,|_{n=0}^{N-1}\}$\\
            \noalign{\vskip 3 pt}
            \hline\noalign{\vskip 3 pt}
  $D^{\scriptscriptstyle(\pm,{\rm even}~Q)}_+$ &
           $Z^n$ & $Z^{n+\nu/2}$ & $Z^{m+N/4}$ &
           $\pm Z^{m-1/2}$ & $\pm Z^{m}$\\
            \noalign{\vskip 3 pt}
  $D^{\scriptscriptstyle(\pm,{\rm even}~Q)}_-$ &
           $Z^n$ & $(-1)^\nu Z^{n+\nu/2}$ & $(-1)^{N/2} Z^{m+N/4}$ &
           $\pm Z^{m-1/2}$ & $\mp Z^{m}$\\
            \noalign{\vskip 3 pt}
  $D^{\scriptscriptstyle(\mu,{\rm even}~Q)}$ & 
           $2Z^n$ & $Z^{n+\nu/2}(t_\mu^\nu+t_\mu^{-\nu})$ &
            $2(-1)^\mu Z^{m+N/4}$ & 0 & 0\\
            \noalign{\vskip 3 pt}
  $D^{\scriptscriptstyle(\bar{\mu},{\rm odd}~Q)}$ & 
           $2Z^n$ & $Z^{n+\nu/2}(t_{\bar{\mu}}^\nu+t_{\bar{\mu}}^{-\nu})$ &
            0 & 0 & 0
\end{tabular}
\label{table4}
\end{table*}

In the following we first review the irreducible representations of
$D_N$ and then discuss how it is modified when $D_N$ is extended
to $ZD_N$. 

Consider first the case with an odd number of islands $N$.  
Using the defining relations (\ref{def_1a})--(\ref{def_1c}) we arrange 
the $2N$ elements of the group $D_N$ into the $(N+3)/2$ conjugacy classes: 
$\{E\}$, $\{T^\nu,T^{-\nu}\}_{\nu=1}^{(N{-}1)/2}$,~and 
$\{T^{m} R\,|_{m=0}^{N-1}\}$.  Firstly, we can construct 
two one-dimensional representations $D^{\scriptscriptstyle(\pm)}$ 
where all operators are represented by numbers.
In these representations, the value of $T$ is determined uniquely:
$D^{\scriptscriptstyle(\pm)}_T=1$, while $R$ may take two values:
$D^{\scriptscriptstyle(\pm)}_R = \pm 1$.  In addition, we can find $(N-1)/2$
representations $D^{\scriptscriptstyle(\mu)}$ with dimensionality two.  The
explicit form of $T$ and $R$ in these representations is:
\begin{equation}
D^{\scriptscriptstyle(\mu)}_T=\pmatrix{t_\mu & 0 \cr 0 & t_\mu^{-1}}, \qquad
D^{\scriptscriptstyle(\mu)}_R=\pmatrix{0 & 1 \cr 1 & 0},
\end{equation}
where the parameter $\mu$ labeling representations takes integer values
$1,\dots,(N-1)/2$, and $t_\mu=\exp(2\pi i \mu/N)$.  We thus have found all the
$(N+3)/2$ irreducible representations; they are listed in the character TABLE
I.

The irreducible representations of the extended dihedral group $ZD_N$ are
found in an analogous way. Assume $N$ odd first. Then $ZD_N$ contains
$N(N+3)/2$ conjugacy classes: $\{Z^n\}_{n=0}^{N-1}$, $\{Z^n
T^\nu,Z^{n+\nu}T^{-\nu}\}_{n=0,\nu=1}^{(N-1),(N-1)/2}$, and
$\{Z^{n-m}T^{2m}R\,|_{m=0}^{N-1}\}_{n=0}^{N-1}$. The construction of the
irreducible representations again follows the scheme described above: for
one-dimensional representations $D^{\scriptscriptstyle(\pm,Q)}$ we find $2N$
solutions: $D^{\scriptscriptstyle(\pm,Q)}_T = \sqrt{Z}$ and
$D^{\scriptscriptstyle(\pm,Q)}_R = \pm 1/\sqrt{Z}$, where we choose that
branch of the square root which puts $\sqrt{Z}$ onto one of the roots
$\exp(2\pi i n/N)$ [for odd $N$ either $\sqrt{Z}$ or $-\sqrt{Z}$ belongs to
the set $\{\exp(2\pi i n/N)|_{n=0}^{N-1}\}$, see Fig.~\ref{reps}].  The
remaining two-dimensional representations $D^{\scriptscriptstyle(\mu,Q)}$ may
again be constructed explicitly:
\begin{equation}
D^{\scriptscriptstyle(\mu,Q)}_T=
             \sqrt{Z}\pmatrix{t_\mu & 0 \cr 0 & t_\mu^{-1}}, \quad
D^{\scriptscriptstyle(\mu,Q)}_R=\pmatrix{0 & 1 \cr Z^{-1} & 0},
\end{equation}
where for $\sqrt{Z}$ we again take one of the roots $\exp(2\pi i n/N)$, the
parameter $\mu$ takes integer values $1,\dots,(N-1)/2$, and $t_\mu=\exp(2\pi i
\mu/N)$.  With $2N$ one-dimensional and $N(N-1)/2$ two-dimensional
representations we then have constructed all irreducible representations of
$ZD_N$.  The result is summarized in TABLE II.

A similar analysis for the qubit loop with an even number $N$ of junctions
produces the character TABLES III and IV for the groups $D_N$ and $ZD_N$
respectively.

In summary, the energy levels of the Josephson junction loop may be classified
according to the irreducible representations of its symmetry group $ZD_N$ at a
given $Q$.  The representation tables (Tables II and IV) look slightly
different for odd and even $N$. All representations are either one- or
two-dimensional. Given an odd $N$, for any value of $Q$, there are two
one-dimensional representations and $(N-1)/2$ two-dimensional representations.
Given an even $N$, for even $Q$ we have four one-dimensional and $(N/2-1)$
two-dimensional representations, while for odd $Q$ there are no
one-dimensional and $N/2$ two-dimensional representations.  The
representations may be distinguished by the eigenvalues of the operator $T$
involved, see Fig.~\ref{reps}.

\begin{figure}[!t]
  \centerline{\epsfxsize = 7.0cm \epsfbox{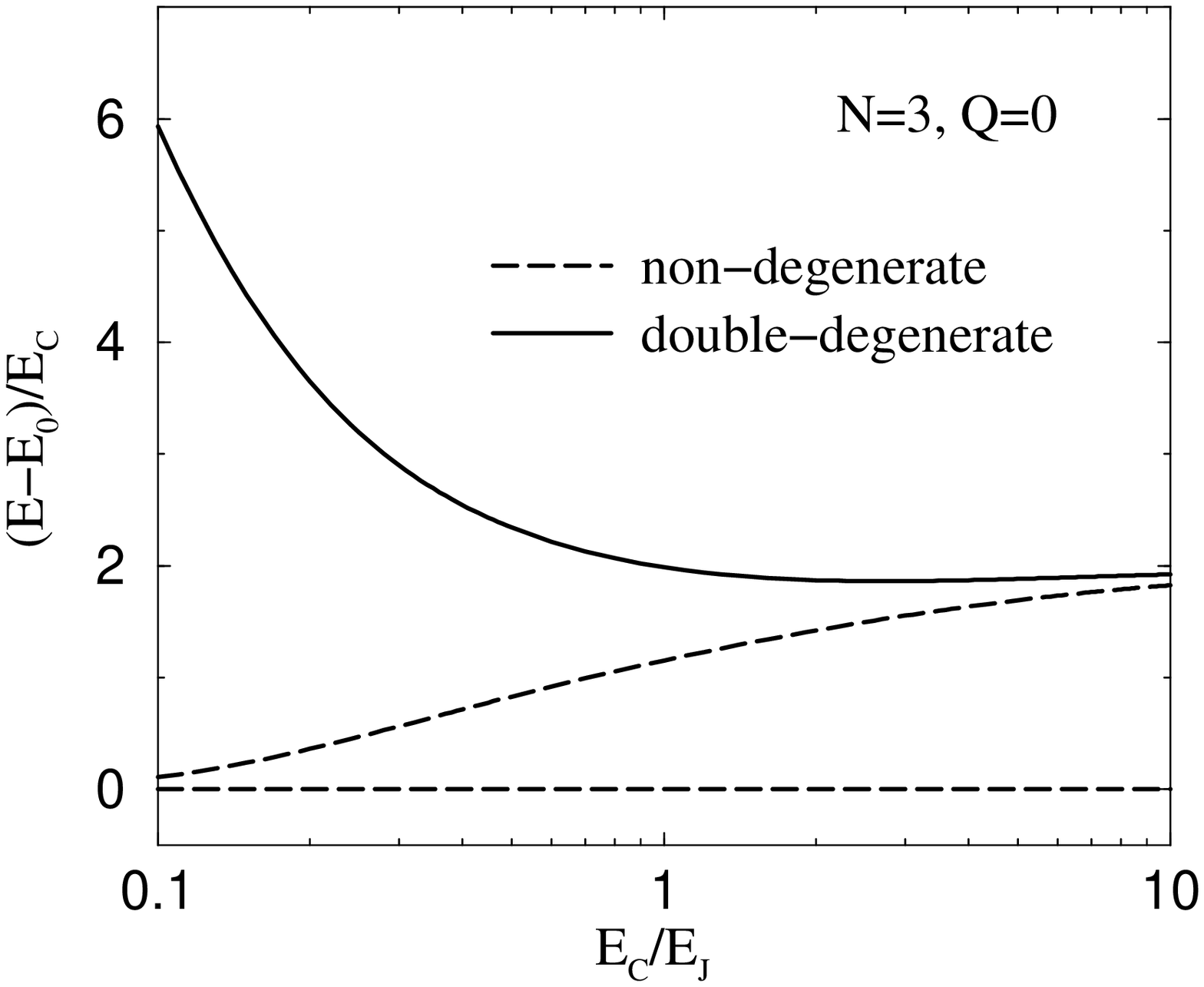}} \centerline{\epsfxsize
    = 7.0cm \epsfbox{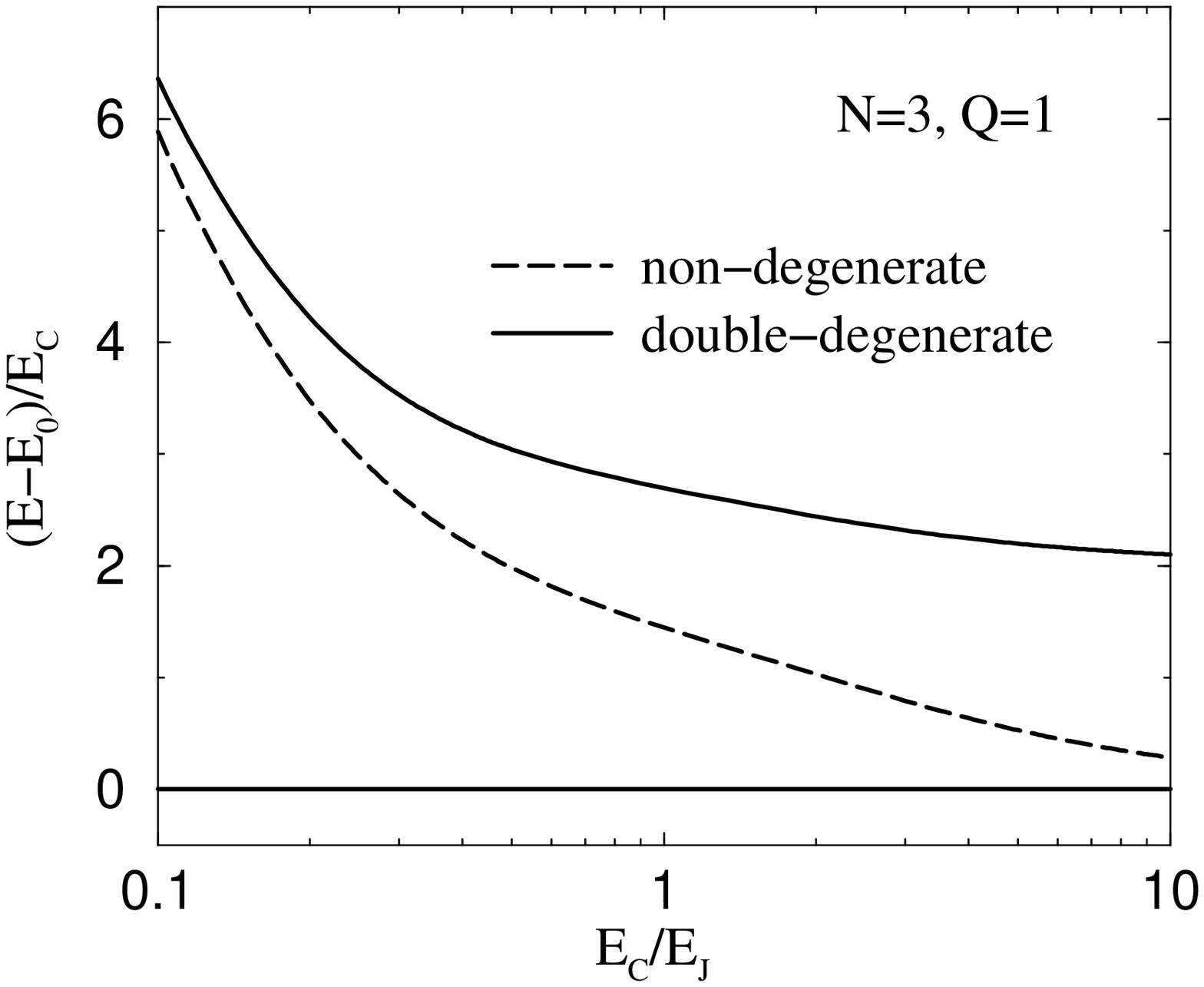}} \narrowtext\hspace{0.5cm} \caption{The two
    lowest excitation energies for the symmetric Josephson junction loop with
    $N=3$ islands, diagonal capacitance matrix $(C^{-1})_{ij}=E_C
    \delta_{ij}$, and vanishing gate voltages $V_i = 0$. The ground-state
    energy at each value of $E_C/E_J$ is subtracted. (a) in the $Q=0$ sector
    where the ground state is non-degenerate; (b) in the $Q=1$ sector with the
    ground state doubly degenerate.  Note the large (small) gap of order $E_C$
    ($E_J/N$) for the insulating state with $Q=0$ (the metallic state with
    $Q=1$) in the limit $E_c/E_J \rightarrow \infty$.}
  \label{spectra}
\end{figure}

Connecting back to our physical problem we can find the degeneracy of the
states: for example, when $N$ is even and $Q$ is odd (see Fig.~\ref{reps}(c))
all states (including the ground state) are doubly degenerate. More work is
needed to determine whether the ground state is degenerate in the other cases.
Inspection of the ground state wave functions in the limits $E_J\gg E_C$ and
$E_J\ll E_C$ shows that they transform with a representation involving the
eigenvalues $T=1$ and $T=\exp(2\pi i Q/N)$. Indeed, in the $E_J \gg E_C$
limit, the state $\{\phi_i=-\pi/N\}$ is a $T=1$ eigenstate, while the state
represented by the points $\{\phi_i=\pi/N,~i\neq k;~\phi_k=\pi/N-2\pi\}$,
$k=1,\dots,N$ by virtue of (\ref{bc}) belongs to the eigenvalue $T=\exp(2\pi i
Q/N)$. In the limit $E_J \ll E_C$, $T$ measures the total momentum of the
repulsive bosons which (for a diagonal capacitance matrix) equals the total
momentum of the tight-binding fermions. The latter is easily read off Fig.\ 
\ref{tight-binding} and accounting for the shift $\pi Q/N$ due to the boundary
condition (originating from the half-quantum flux) we again obtain the two
eigenvalues $T=1$ and $T=\exp(2\pi i Q/N)$ for the two lowest states. Thus if
$Q \neq 0~({\rm mod}~N)$ the two lowest states combine into a two-dimensional
representation (and the same in both limits), resulting in a doubly degenerate
ground state. For $Q$ divisible by $N$, the ground state corresponds to a
one-dimensional representation of $ZD_N$, i.e., it is non-degenerate. As the
ratio $E_J/E_C$ is swept from one limit to the other, the ground state
continuously evolves preserving its degeneracy, unless a level-crossing occurs
with a level of different symmetry.  We have confirmed numerically that such a
level crossing does not occur for the cases $N=3$ and $N=4$ using a diagonal
capacitance matrix $(C^{-1})_{ij}$, (the results for $N=3$ are shown in
Fig.~\ref{spectra}), and we believe that this property holds for any $N \ge
3$.

\section{Requirements for interference observation}

The experimental observation of the quantum interference effect discussed in
this paper relies on several requirements on the Josephson junction loop.
First, we have to assume that all quasi-particles are frozen out, which
defines an upper bound on the temperature roughly estimated as $T <
T^*\sim\Delta_{\rm sc}/\log (\nu_0 V \Delta_{\rm sc})$, where $\nu_0$ is the
density of states at the Fermi level, $\Delta_{\rm sc}$ is the superconducting
gap, and $V$ is the volume of the system~\cite{Tuominen}.  For a
micrometer-size aluminium loop similar to that used in the experimental work
of van der Wal {\it et al.}~\cite{Mooij2}, $T^* \sim \Delta_{\rm sc}/15 \sim
100~{\rm mK}$ and can be easily achieved.

Second, we assume that no charge tunneling is possible onto the array.  In
practice, we may allow the array to be connected to an external reservoir via
a large resistor (to be able to change the charge on the array by an overall
shift of gate voltages). Its resistance then must be much larger than the
characteristic resistance scale $R^*\sim (C \Delta)^{-1}$, where $C$ is the
capacitance of the array, and $\Delta$ is the relevant energy scale (of the
order of the splitting between the two lowest states). In the charge-dominated
limit, $\Delta \sim E_C$ and $R^*$ is of the order of the resistance quantum
$h/4e^2$, in agreement with the conventional condition for charge quantization
in the Coulomb blockade setup\cite{AverinLikharev}. In the phase-dominated
limit, $\Delta \ll E_C$ and $R^*$ is much larger than the resistance quantum.
Also, under the condition $\Delta_{\rm sc} > E_C^{\rm loop}$ a `weak' contact
to the external world would allow an unpaired quasi-particle to escape from
the loop, leaving only paired electrons in the system.

Third, the islands and the junctions are assumed to be identical. The
precision to which this symmetry has to hold in the phase-dominated limit may
be simply estimated from the condition that the tunneling actions agree up to
small deviations of order one. The tunneling action scales as $S \sim
(E_J/E_C)^{1/2}$ and therefore the required precision is $(\delta E_J/E_J),
(\delta E_C/E_C) \ll S^{-1}$. For a detectable splitting $S$ must be not too
large; for a qubit design, $S$ is typically\cite{Levitov,Blatter-2} taken to
be of order 5--10, and these conditions may well be satisfied.

The required precision on the magnitude of the external magnetic field (or,
equivalently, on the flux through the loop) may be estimated from the
condition that the level splitting due to the deviation of the flux from
$\Phi_0/2$ is much smaller than that from the tunneling between the qubit
states. In the phase-dominated limit ($E_J\gg E_C$) this produces the
condition $E_J \delta\Phi/\Phi_0 \ll \Delta$ and hence $\delta\Phi/\Phi_0 \ll
S^{-1} e^{-S}$. In the opposite charge-dominated limit, it is sufficient to
require that $\delta\Phi/\Phi_0 \ll 1$.

In conclusion we have discussed the spectral properties of symmetric Josephson
junction loops, devices similar to those recently proposed as potential qubits
for quantum computing\cite{Levitov,Blatter-2}.  While in the charge-dominated
regime with $E_C\gg E_J$ the dependence of the ground-state degeneracy on the
total charge on the island is a simple charging effect, this degeneracy
derives from a subtle quantum interference in the phase-dominated limit
$E_J\gg E_C$. Several proposals on solid state realizations of quantum bits
belong to the latter limit. From our analysis it follows that the requirements
for observing the charge dependence of the qubit-level splitting are similar
to those for the qubit operation (plus symmetry requirements which are easy to
satisfy). The interference effects studied in this paper then may serve as a
good test of quantum coherence in such qubit designs.

We thank M.~Feigelman and O.~Hallatschek
for helpful discussions, and M.~Troyer for 
providing the exact diagonalization code. We thank the 
Swiss National Fonds for financial support. L.~I.
thanks ETH Z\"urich for hospitality.



\begin{references}
  
\bibitem{JJA} R.S.~Newrock, C.J.~Lobb, U.~Geigenm\"uller, and M.~Octavio,
  ``The two-dimensional physics of Josephson junction arrays'', Solid State
  Physics 54 (2000), 263, and references therein.

\bibitem{Blanter}
Ya.M.~Blanter, R.~Fazio, and G.~Sch\"on, ``Duality in Josephson junction
arrays'', cond-mat/9701223 and in Nucl.~Phys.~B.

\bibitem{Aharonov-Casher}
Y.~Aharonov and A.~Casher,
``Topological quantum effects for neutral particles'',
Phys.~Rev.~Lett. 53 (1984), 319.

\bibitem{Mooij3}
W.J.~Elion, J.J.~Wachters, L.L.~Sohn, and J.E.~Mooij, ``Observation of the
Aharonov-Casher effect for vortices in Josephson-junction arrays'',
Phys.~Rev.~Lett. 71 (1993), 2311.

\bibitem{Schoen} A.~Shnirman, G.~Sch\"on, and Z.~Hermon,
``Quantum manipulations of small Josephson junctions'',
Phys.~Rev.~Lett.  79 (1997), 2371.

\bibitem{Levitov}
T.P.~Orlando, J.E.~Mooij, L.~Tian, C.H.~van~der~Wal, L.S.~Levitov, S.~Lloyd,
and J.J.~Mazo, ``Superconducting persistent-current qubit'', Phys.~Rev. B 60
(1999), 15398.

\bibitem{Blatter}
L.~B. Ioffe, V.~B. Geshkenbein, M.~V. Feigelman, 
A.L. Fauch\`ere, and G.~Blatter, ``Environmentally 
decoupled sds-wave Josephson junctions for quantum computing'',
Nature 398 (1999), 679.

\bibitem{Blatter-2}
G.~Blatter, V.B.~Geshkenbein, and L.B.~Ioffe, ``Engineering superconducting
phase qubits'', e-print cond-mat/9912163.

\bibitem{central-extension}
See e.g.\ A.O.~Barut and R.~Raszka, ``Theory of group representations and
applications'', World Scientific, Singapore 1986.

\bibitem{Mooij1}
C.H.~van~der~Wal and J.E.~Mooij, ``Controlled single-Cooper-pair charging
effects in a small Josephson junction array'', J.~Superconduct. 12 (1999), 807.

\bibitem{Mooij2}
C.H.~van~der~Wal, A.C.J.~ter~Haar, F.K.~Wilhelm, R.N. Schouten,
C.J.P.M.~Harmans, T.P.~Orlando, S.~Lloyd, and J.E.~Mooij, ``Quantum
superposition of macroscopic persistent-current states'', Science 290 (2000),
773.

\bibitem{Ivanov}
D.A.~Ivanov and M.V.~Feigel'man, ``Coulomb effects in a ballistic one-channel
S-S-S device'', Zh.~Eksp.~Teor.~Fiz. 114 (1998), 640 [J.~Exp.~Theor.~Phys. 87
(1998), 349; e-print cond-mat/9712074].

\bibitem{Mooij-charge}
P.~Lafarge, M.~Matters, and J.E.~Mooij, ``Charge representation of a small
two-dimensional Josephson-junction array in the quantum regime'', Phys.~Rev. B
54 (1996), 7380.

\bibitem{Pokrovsky}
V.A.~Kalatsky and V.L.~Pokrovsky, ``Spectra and magnetic properties of
large spins in external fields'', Phys.~Rev. A 60 (1999), 1824.

\bibitem{Tuominen}
M.T. Tuominen, J.M. Hergenrother, T.S. Tighe, and M.~Tinkham,
``Experimental evidence for parity-based $2e$ periodicity in
a superconducting single-electron tunneling transistor'',
Phys.~Rev.~Lett. 69 (1992), 1997.

\bibitem{AverinLikharev} D.V.\ Averin and K.K.\ Likharev,
in {\it Mesoscopic Phenomena in Solids}, edited by B.\ Altshuler
{\it et al.}~(Elsevier, Amsterdam, 1991).

\end{references}
\end{document}